\documentclass[aip,reprint]{revtex4-1}
\usepackage{graphicx}

\draft 

\begin{document}


\title{Path Integral Molecular Dynamics for Fermions: Alleviating the Sign Problem with the Bogoliubov Inequality} 



\author{Barak Hirshberg}
 \email{barakh@ethz.ch}
\affiliation{Department of Chemistry and Applied Biosciences, ETH Zurich, 8092 Zurich, Switzerland}
\affiliation{Institute of Computational Sciences, Universit\`{a} della Svizzera italiana, via G. Buffi 13, 6900 Lugano, Switzerland}%

\author{Michele Invernizzi}%
\affiliation{Institute of Computational Sciences, Universit\`{a} della Svizzera italiana, via G. Buffi 13, 6900 Lugano, Switzerland}%
\affiliation{National Centre for Computational Design and Discovery of Novel Materials MARVEL, Universit\`{a} della Svizzera italiana, via G. Buffi 13, 6900 Lugano, Switzerland}%
 \affiliation{Department of Physics, ETH Zurich, 8092 Zurich, Switzerland}

\author{Michele Parrinello}
 \affiliation{Department of Chemistry and Applied Biosciences, ETH Zurich, 8092 Zurich, Switzerland}
\affiliation{Institute of Computational Sciences, Universit\`{a} della Svizzera italiana, via G. Buffi 13, 6900 Lugano, Switzerland}%
\affiliation{Atomistic Simulations, Italian Institute of Technology, Via Morego 30, 16163 Genova, Italy}



\begin{abstract}
We present a method for performing path integral molecular dynamics (PIMD) simulations for fermions and address its sign problem. 
PIMD simulations are widely used for studying many-body quantum systems at thermal equilibrium. However, they assume that the particles are distinguishable and neglect bosonic and fermionic exchange effects. Interacting fermions play a key role in many chemical and physical systems, such as electrons in quantum dots and ultracold trapped atoms. A direct sampling of the fermionic partition function is impossible using PIMD since its integrand is not positive definite. We show that PIMD simulations for fermions are feasible by employing our recently developed method for bosonic PIMD and reweighting the results to obtain fermionic expectation values. The approach is tested against path integral Monte Carlo (PIMC) simulations for up to 7 electrons in a two-dimensional quantum dot for a range of interaction strengths. However, like PIMC, the method suffers from the sign problem at low temperatures. We propose a simple approach for alleviating it by simulating an auxiliary system with a larger average sign and obtaining an upper bound to the energy of the original system using the Bogoliubov inequality.  This allows fermions to be studied at temperatures lower than would otherwise have been feasible using PIMD, as demonstrated in the case of a three-electron quantum dot. Our results extend the boundaries of PIMD simulations of fermions and will hopefully stimulate the development of new approaches for tackling the sign problem.
\end{abstract}


\maketitle 

\section{Introduction}
\label{sec:Intro}
Interacting fermions at finite temperature are fundamental to a wide range of chemical and physical phenomena. Examples include ultracold trapped atoms~\cite{Bloch2005, Haller2015, Schneider2012}, electrons in quantum dots~\cite{Ellenberger2006, Yannouleas2007, Li2007} and warm dense matter~\cite{Cytter2019}. Developing simulation methods that accurately describe correlated fermions at various thermodynamic conditions is therefore very desirable~\cite{Wagner2016}. 
The path integral (PI) formulation of quantum mechanics~\cite{Feynman2005} is a powerful approach for studying many-body systems at finite temperature. It is based on the observation that the partition function of a quantum system is isomorphic to the partition function of a fictitious, extended classical system~\cite{Chandler1981}. Assuming the quantum particles are distinguishable, each one is represented by a classical ``ring polymer" composed of $P$ replicas of the particle (the ``beads" of the ring polymer) connected by harmonic springs~\cite{Markland2018}. The frequency of the springs is proportional to the temperature and number of beads. The exact quantum result is obtained in the limit $P \rightarrow \infty $. Beads having the same index but representing different particles interact through a scaled interaction potential. The partition function of the fictitious classical system is typically sampled using molecular dynamics (MD)~\cite{Parrinello1984, Markland2018} or Monte Carlo (MC) methods~\cite{Ceperley1995}.

\begin{figure}
    \centering
    \includegraphics[width=1.0\columnwidth]{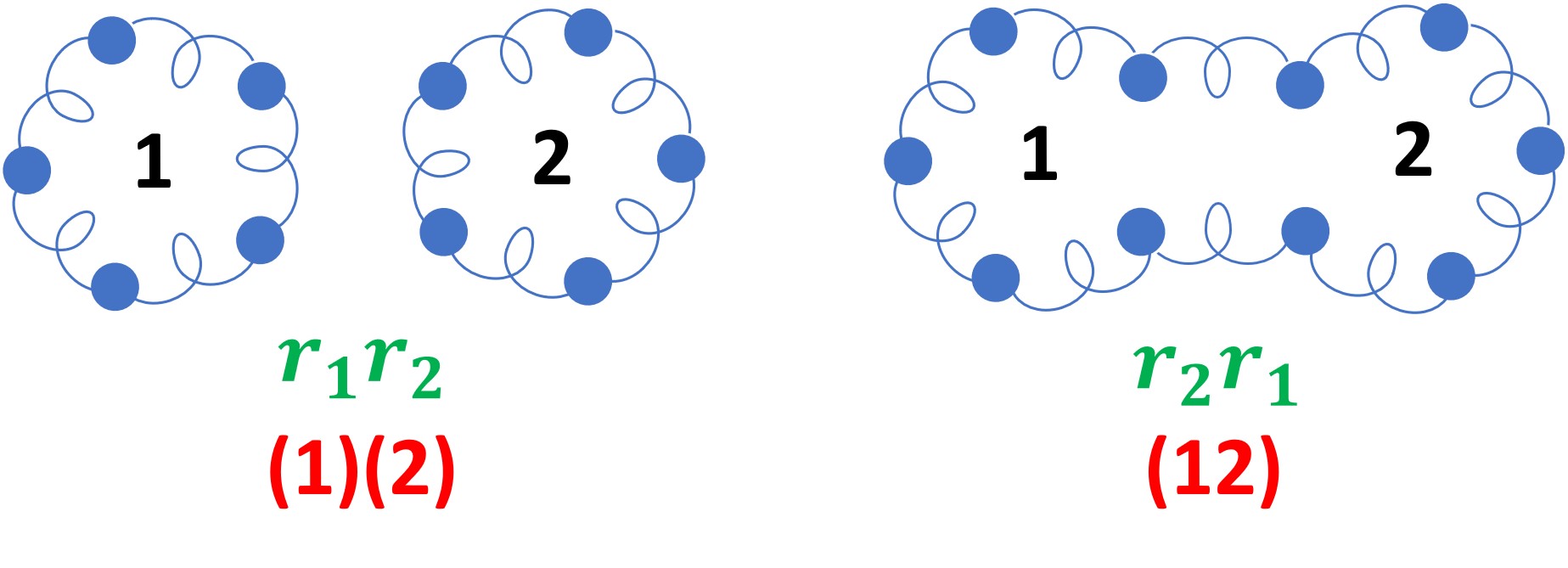}
    \caption{Ring polymer configurations for two indistinguishable particles. The permutations of two particles are shown in green and the corresponding cycle notations are shown in red. The ring polymer configurations can be directly inferred from the cycle notation for permutations.}
    \label{fig:Perm}
\end{figure}

For $N$ indistinguishable particles, bosons or fermions, one must consider all $N!$ particle permutations. As a result, the fictitious classical system is no longer composed of a single ring polymer for each particle. Ring polymer configurations in which particles are connected into longer rings must be included~\cite{Ceperley1995,Lyubartsev1993}, as shown in Fig.~\ref{fig:Perm} in the two-particle case. The number of ring polymer configurations scales exponentially with system size and enumerating them is impractical for more than a few atoms. In Path Integral Monte Carlo (PIMC) simulations this problem is avoided by using MC moves designed to sample particle permutations, as pioneered by Ceperley and coworkers~\cite{Ceperley1995, Pollock1984, Ceperley1986, Pollock1987, Ceperley1992}. 
Recently, we showed that the potential and forces required for path integral molecular dynamics (PIMD) simulations for bosons can be evaluated recursively without enumerating or sampling particle permutations~\cite{Hirshberg2019}. The resulting algorithm scales cubically with system size, which has allowed the first PIMD simulations of large bosonic systems to be performed. In this Communication, we show how this method can be extended to obtain thermodynamic properties for fermionic systems.

In comparison to bosons, fermions present an additional challenge since even and odd sequential particle permutations carry a positive and negative sign, respectively. As a result, direct sampling of fermionic partition functions is problematic, either in MC or MD, since their integrands are not positive definite. One approach to overcome this problem within the framework of PIMC is the fixed node approximation, often referred to as restricted PIMC~\cite{Ceperley1992}. But since the exact nodal surface of the fermionic density matrix is usually unknown, an approximate ansatz is employed, such as the free electrons density matrix~\cite{Brown2013}. 
An alternative approach is to simulate the system as if it were composed of bosons and reweight the results according to the sign of each sampled configuration~\cite{Takahashi1984}. However, this procedure leads to the infamous fermionic sign problem - the exponentially slow convergence of expectation values due to delicate cancellation of positive and negative contributions~\cite{DuBois2017}. It was shown by Troyer and Wiese that a general solution to the sign problem  - an algorithm which converges in polynomial time - is ``almost certainly unattainable''~\cite{Troyer2005}. Even if a complete solution is unlikely, it is highly desirable to develop new methods for simulating fermions which may provide avenues for alleviating the sign problem. 

Recently, several methods have been proposed to alleviate the sign problem within PIMC, such as configuration PIMC~\cite{Schoof2011} and permutation blocking PIMC~\cite{Dornheim2015}. 
In contrast, only a few methods have been proposed to perform fermionic PIMD simulations in general and even fewer studies have addressed the sign problem in this context. Miura and Okazaki proposed to sample the absolute value of the probability density for fermions~\cite{Miura2000}. The potential is obtained from the absolute value of a determinant, which accounts for all particle permutations.  Expectation values for fermions are obtained by reweighting the contribution of different configurations according to the sign of their determinant. However, this method was applied to only three particles and its sign problem was not investigated in detail. Another limitation of this approach is that the potential has an infinite barrier at the nodes of the fermionic probability density, effectively preventing the system from exploring all regions of phase space. 
More recently, Runeson et al.~\cite{Runeson2018} have shown that by combining enhanced sampling techniques with PIMD simulations of distinguishable particles, the problem of two electrons in a quantum dot can be solved. However, it is not clear how to extend their work to larger systems.

In this Communication, we propose a different approach to performing PIMD simulations of fermions. We employ our recently developed method for bosonic PIMD~\cite{Hirshberg2019} and reweight the results to obtain fermionic expectation values. The method is applicable to any system for which the bosonic simulation can be performed and naturally avoids the problem of the infinite barriers, since the bosonic probability density is nodeless. However, as in PIMC, the method suffers from the sign problem at low temperatures. We propose a simple approach to alleviate the sign problem based on the Bugoliubov inequality for the free energy~\cite{feynman1972}. This allows us to estimate the energies of fermionic systems at lower temperatures than would otherwise have been possible using PIMD. In the following, we first present the method in detail. Then, we benchmark the new approach against recent fermionic PIMC simulations by Dornheim~\cite{Dornheim2019} for up to $N=7$ electrons in a two-dimensional quantum dot (QD). In Section~\ref{sec:sign}, we propose to use the Bogoliubov inequality to alleviate the sign problem and apply it to a three-electron QDs at low temperatures. The final section presents the conclusions and briefly discusses future directions.

\section{Theory}
\label{sec:Theory}

The Hamiltonian operator for a system of $N$ identical particles of mass $m$, interacting through a potential $\hat{V}$ is

\begin{equation}
\hat{H} = \frac{1}{2m} \sum_{l=1}^N {\bf \hat{p}}^2_l + \hat{V}({\bf r}_1,...,{\bf r}_N).
\label{eq:H}
\end{equation}

The partition function is obtained by taking the trace of the density operator for a thermal state, $Z = Tr\{e^{-\beta \hat{H}}\}$, where $\beta = (k_B T)^{-1}$ is the inverse temperature and $k_B$ is the Boltzmann constant. The path integral expression for the partition function\cite{tuckerman2010} is





\begin{equation}
Z_I \sim \lim_{P\to\infty} \int e^{-\beta U^{(N)}_I} dR_1 ... dR_N,
\label{eq:Z_PI}
\end{equation}
where $I$ represents whether the particles are bosons (B), fermions (F) or distinguishable (D) and $R_l$ represents collectively the coordinates of $P$ replicas of particle $l$,  $ {\bf r}_l^1,..., {\bf r}_l^P$. In practice, $P$ is increased until the desired expectation values are converged. Here we focus on sampling the partition function of Eq.~\ref{eq:Z_PI} using MD simulations.

For distinguishable particles, the potential is given by

\begin{equation}
U_D^{(N)} = \frac{1}{2}m \omega_P^2 \sum_{l=1}^N \sum_{j=1}^P ({\bf r}_l^{j+1} - {\bf r}_l^j)^2 + \frac{1}{P}\sum_{j=1}^P V({\bf r}_1^j,...,{\bf r}_N^j),
\label{eq:VD_PI}
\end{equation}
where $\omega_P=\sqrt{P}/\beta \hbar$ and ${\bf r}_l^{P+1} = {\bf r}_l^1$.  The first and second terms arise from the kinetic and potential energy operators of the quantum system, respectively.
Eq.~\ref{eq:VD_PI} is isomorphic to the partition function of $N$ classical ring polymers, each one composed of $P$ beads which are connected by harmonic springs\cite{Chandler1981}. 
Beads $j$ of different ring polymers (representing different quantum particles) interact through a scaled interaction potential.

In order to treat bosons and fermions, the trace should be taken over a properly symmetrized or anti-symmetrized basis, respectively.
Since the potential energy is invariant under permutations of identical particles, only the term arising from the kinetic energy operator is affected.
Each permutation leads to a ring polymer configuration which can be directly inferred from the cycle notation for the permutation, as demonstrated in Fig.~\ref{fig:Perm} for two particles.
Since the number of permutations is $N!$, direct enumeration of all ring-polymer configurations is possible only for very small systems. 
We have recently shown that the potential for bosons can be evaluated recursively, avoiding the need to enumerate the exponentially large number of permutations~\cite{Hirshberg2019}.
The algorithm scales cubically with system size, which allowed the first applications of PIMD to large bosonic systems.
It is also possible to combine PIMC for bosons with PIMD for distinguishable particles in a hybrid algorithm~\cite{Walewski2014,Walewski2014a,Uhl2018,Schran2018,Uhl2019} to treat large systems. However, our method is simpler as it treats both distinguishable and indistinguishable particles on equal footing.

Extending our bosonic PIMD approach~\cite{Hirshberg2019} to fermions is achieved by defining the potential as 
\begin{equation}
U_{B/F}^{(N)} = -\frac{1}{\beta} \ln W_{B/F}^{(N)} + \frac{1}{P}\sum_{j=1}^P V({\bf r}_1^j,...,{\bf r}_N^j), 
\label{eq:VBorF_PI}
\end{equation}
and evaluating the argument of the logarithm by using the recurrence relation

\begin{equation}
W_{B/F}^{(N)} = \frac{1}{N} \sum_{k=1}^N \xi^{k-1} e^{-\beta E_N^{(k)}} W_{B/F}^{(N-k)}.
\label{eq:WBorF_PI}
\end{equation}
In Eq.~\ref{eq:WBorF_PI}, $\xi=1$ for bosons, $\xi=-1$ for fermions and we set $W_{B/F}^{(0)}=1$ .
$E_N^{(k)}(R_{N-k+1},...,R_N) $ is the spring energy of k particles connected together in a long ring, given by 

\begin{equation}
E_N^{(k)} = \frac{1}{2}m \omega_P^2 \sum_{l=N-k+1}^N \sum_{j=1}^P \left( {\bf r}_l^{j+1} - {\bf r}_l^{j} \right) ^2.
\label{eq:Ek}
\end{equation}
In Eq.~\ref{eq:Ek}, $r_l^{P+1}=r_{l+1}^1$ except for $l=N$, for which $r_N^{P+1}=r_{N-k+1}^1$.
We note that Eq.~\ref{eq:VBorF_PI} and~\ref{eq:WBorF_PI} lead to a slightly different expression for the potential for fermions than the one obtained by evaluating a determinant to account for particle permutations~\cite{Takahashi1984, Miura2000}. However, both methods converge to the same quantum-mechanical expectation values after integration over all particles in the partition function.
Expectation values are evaluated using the regular estimators. For a position dependent operator $\hat{O}({\bf r}_1,..,{\bf r}_N)$, such as the potential energy, the estimator $ \varepsilon_O $ is

\begin{equation}
\varepsilon_O = \frac{1}{P}\sum_{j=1}^P O({\bf r}_1^j,..,{\bf r}_N^j),
\label{eq:Est}
\end{equation}
and the expectation value is obtained by 

\begin{equation}
\left\langle \hat{O} \right\rangle_I = \frac{1}{Z_I} \int \varepsilon_O e^{-\beta U^{(N)}_I} dR_1 ... dR_N \equiv \left\langle \varepsilon_O \right\rangle_I.
\label{eq:Av}
\end{equation}
For evaluating the kinetic energy we use the virial estimator~\cite{Herman1982} (see SI for more details).

The key difference between fermions and bosons is that, in the case of fermions, odd permutations carry a negative sign. Therefore, $W_F^{(N)}$ is no longer positive definite. When it becomes negative the potential $U_F^{(N)}$ is complex and cannot be sampled in standard MD or MC simulations. As mentioned in Section~\ref{sec:Intro}, using the absolute value of $W_F^{(N)}$ will result in a real potential, but with an infinite barrier when $W_F^{(N)}=0$. 

Here, we propose instead to simulate the system as if it were composed of bosons, similarly to what is done in PIMC simulations~\cite{Troyer2005}.  
The fermionic expectation values are then recovered from the bosonic simulations by reweighting

\begin{equation}
\left\langle \hat{O} \right\rangle_F = \frac{\left\langle \varepsilon_O s \right\rangle_B}{\left\langle s \right\rangle_B},
\label{eq:reW}
\end{equation}
where $s = W_F^{(N)}/W_B^{(N)}$ is the signed relative weight for each sampled configuration. The evaluation of $W_F^{(N)}$ is done concurrently with the evaluation of the bosonic potential, and therefore presents only a modest additional computational cost. 
The average sign $ \langle s \rangle_B $ decreases exponentially with $\beta$ and the number of particles~\cite{Troyer2005}. As a result, the relative error for expectation values evaluated using Eq.~\ref{eq:reW} becomes exponentially hard to converge, which is referred to as the fermionic sign problem. In the next Section, we show that as long as  $\langle s \rangle_B$ is large enough, the simulations of electrons in two-dimensional QDs can be converged and are in very good agreement with PIMC results~\cite{Dornheim2019}. Then, we propose a procedure to alleviate the sign problem and evaluate the thermal energy of fermionic systems at lower temperatures than would otherwise have been possible using PIMD.

\section{Benchmark}
\label{sec:QDs}

To test the new method for fermionic PIMD we performed simulations of electrons in two-dimensional QDs. These materials have received much attention in recent years due to their interesting electronic, optical and catalytic properties and their potential use in technological applications~\cite{Wang2016, Xu2018, Manikandan2019}.
Landman and collaborators~\cite{Ellenberger2006, Yannouleas2007, Li2007} have shown that the two-dimensional QDs can be modelled using the dimensionless Hamiltonian 

\begin{equation}
\hat{H} = -\frac{1}{2} \sum_{l=1}^N {\bf \nabla}_l^2 + \frac{1}{2}\sum_{l=1}^N {\bf r}_l^2 + \sum_{l,m>l}^N \frac{\lambda}{|{\bf r}_l - {\bf r}_m|}.
\label{eq:H_QD}
\end{equation}

In Eq.~\ref{eq:H_QD}, $ \lambda = e^2 / (\kappa l_0 \hbar \omega_0)$ is the Wigner parameter, $\kappa$ is the relative QD dielectric constant and $\omega_0$ is the frequency of the two-dimensional trap. We denote the effective mass of the electrons as $m$ and $l_0 = \sqrt{\hbar/m \omega_0}$ is the characteristic length of the dot. The Wigner parameter represents the ratio between the screened Coulomb repulsion in the QD and the confinement of the harmonic trap. When $ \lambda < 1 $ exchange effects dominate since the Coulomb repulsion is screened and, conversely, when $ \lambda > 1 $ the repulsive interaction is dominant. In the simulations below, the trap is isotropic with a frequency $\hbar \omega_0 = 5.1$ meV and the effective mass was set to $m= 0.07 m_e$ which are realistic values  for an electron in a GaAs QD~\cite{Ellenberger2006}. We use a development version of LAMMPS~\cite{Plimpton1995} to perform all simulations with a time step of 1 fs and the Nose-Hoover chains thermostat~\cite{Martyna1992}. Expectation values were converged with respect to the number of beads using $P=12$ at most.

\begin{figure}
\includegraphics[width=1.0\columnwidth]{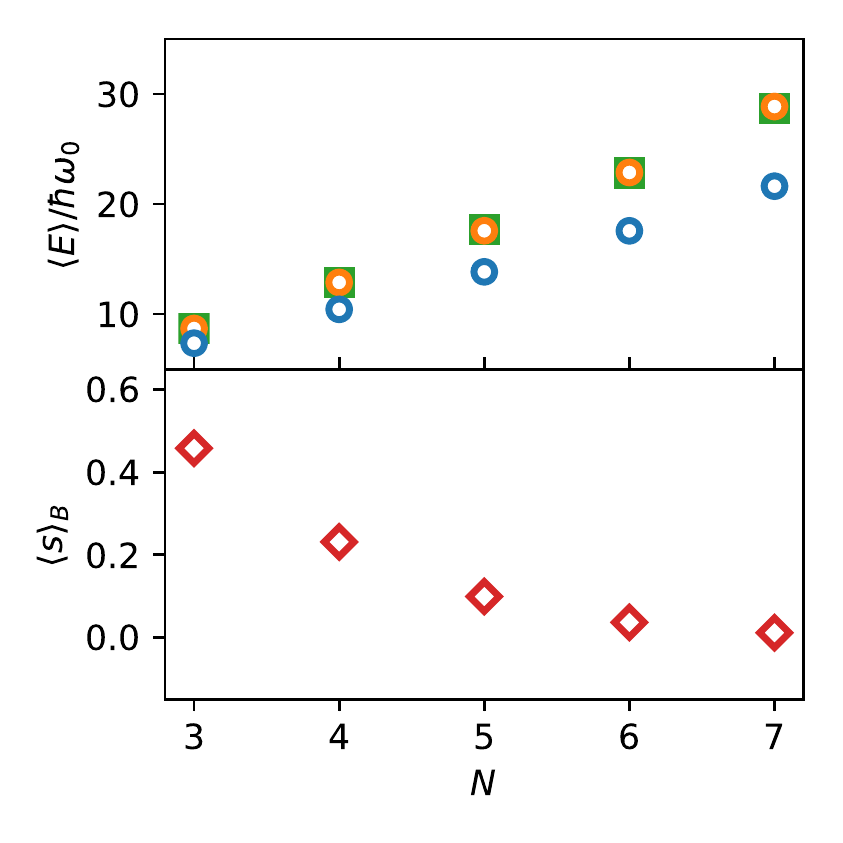}
\caption{\label{fig:EvsN} The average thermal energy (upper panel) and sign (lower panel) as a function of the number of interacting fermions $N$ in a two-dimensional QD. The interaction strength is $\lambda=0.5$ and $\beta \hbar \omega_0=1$. Blue and orange circles represent the results obtained using PIMD for bosons and fermions, respectively. Green squares represent PIMC results of Dornheim~\cite{Dornheim2019}. Error bars are smaller than symbol size.}
\end{figure}

\begin{figure}
\includegraphics[width=1.0\columnwidth]{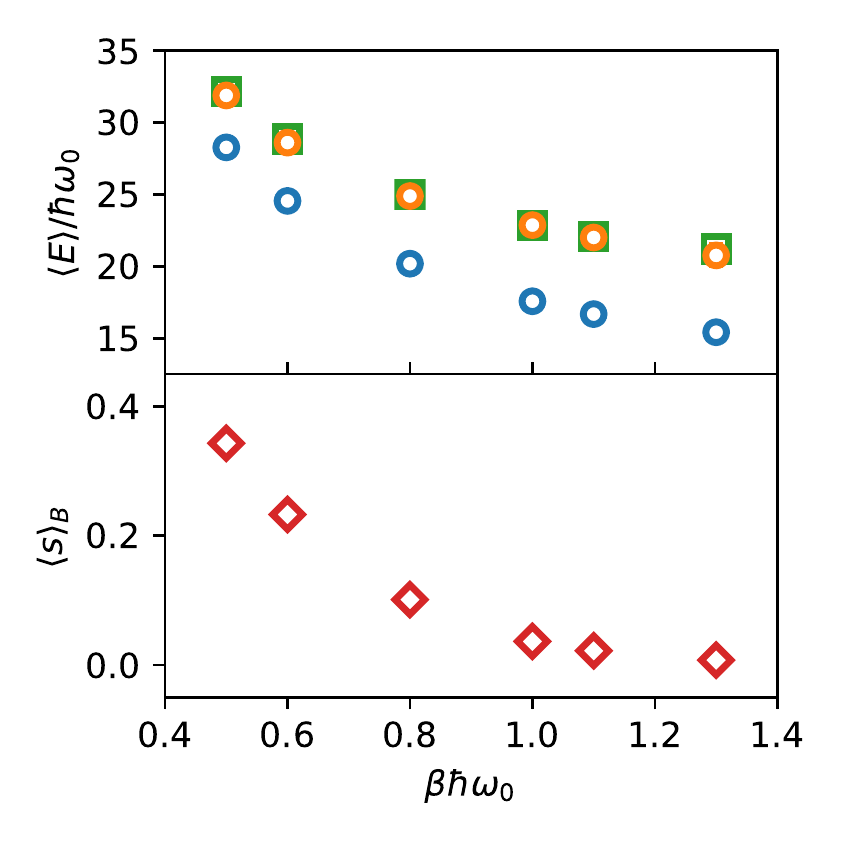}
\caption{\label{fig:Evsbeta} The average thermal energy (upper panel) and sign (lower panel) for $N=6$ interacting fermions ($\lambda=0.5)$ in a two-dimensional QD as a function of inverse temperature. Blue and orange circles represent the results obtained using PIMD for bosons and fermions, respectively. Green squares represent PIMC results of Dornheim~\cite{Dornheim2019}. Error bars are smaller than symbol size.}
\end{figure}

\begin{figure}
\includegraphics[width=1.0\columnwidth]{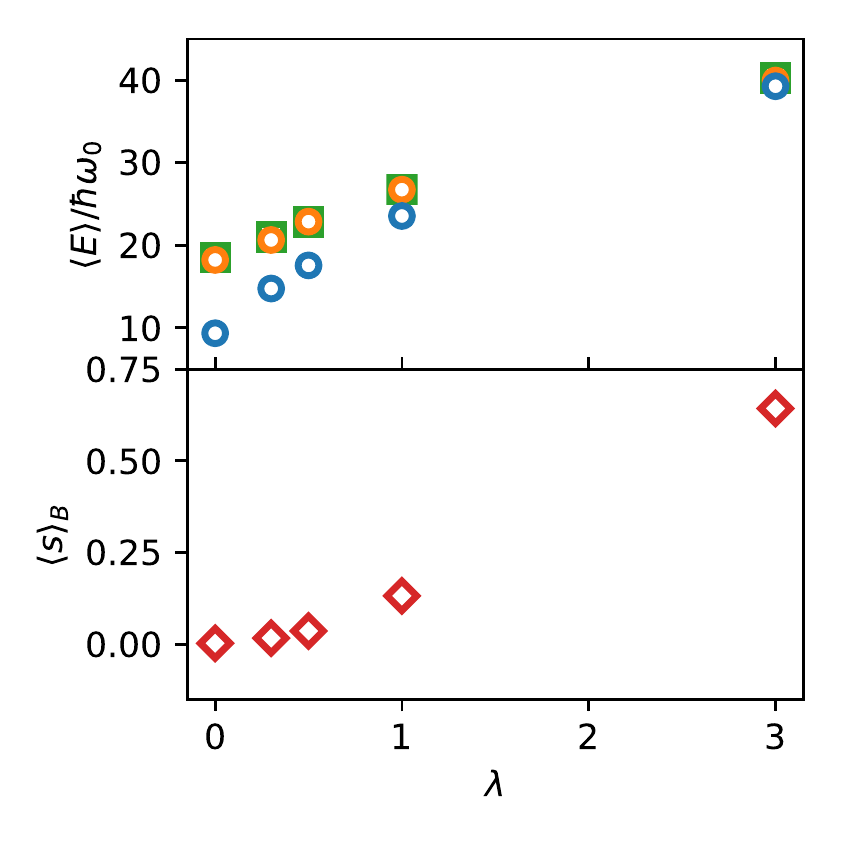}
\caption{\label{fig:Evslambda} The average thermal energy (upper panel) and sign (lower panel) at $\beta \hbar \omega_0=1$ for $N=6$ fermions in a two-dimensional QD as a function of interaction strength $\lambda$. Blue and orange circles represent the results obtained using PIMD for bosons and fermions, respectively. Green squares represent PIMC results of Dornheim~\cite{Dornheim2019}. Error bars are smaller than symbol size.}
\end{figure}

Fig.~\ref{fig:EvsN} shows the average thermal energy for a system with $\beta\hbar\omega_0=1$ and $\lambda=0.5$, as a function of system size. 
Excellent agreement is obtained in comparison to PIMC, with a maximum absolute error of 0.6\% and mean absolute error of 0.4\%.
Fig.~\ref{fig:Evsbeta} and~\ref{fig:Evslambda} present the average thermal energies for $N=6$ as a function of the inverse temperature and repulsion strength. For the range of temperatures ($0.5 \le \beta \le 1$; $\lambda = 0.5$) and interaction parameters ($0 \le \lambda \le 4$; $\beta = 1$) considered here, we obtain very good agreement with PIMC results, with mean absolute errors of 0.8\% and 0.9\%, respectively. The simulations span a realistic range of interaction strength as the value of the interaction parameter in a GaAs QD is $\lambda \approx 1.55 $~\cite{Ellenberger2006}.

The lower panels of Fig.~\ref{fig:EvsN}-\ref{fig:Evslambda} also present the average sign as a function of system size, temperature and repulsion strength. They show that the average sign decreases with increasing $beta$ and number of particles, but also that the sign problem is more severe for weakly interacting than for strongly interacting electrons. This is in agreement with the detailed investigation by Dornheim~\cite{Dornheim2019} and physical intuition.
Converging the fermionic expectation values becomes harder with decreasing average sign. Of the results presented in this Section, the simulations for $N=6$, $\beta=1$ and $\lambda=0$ had the smallest average sign with $\langle s \rangle _B \approx 0.003$. They required 10 independent trajectories of $9\cdot 10 ^8$ MD steps in order to converge the energy to within 2\%. In comparison, the same simulations with $\lambda=3$ ($\langle s \rangle _B \approx 0.64$) required only 5 simulations of $75 \cdot 10 ^6$ MD steps in order to converge the energy to within 0.05\%. The error bars on expectation values were evaluated using a weighted average over the independent simulations and full details are given in the SI. The following section describes the use of the Bogoliubov free energy principle to obtain expectation values for systems with an even smaller average sign.

\section{Alleviating the sign problem}
\label{sec:sign}

Overcoming the sign problem is a formidable challenge. As mentioned above, a general solution is probably unattainable~\cite{Troyer2005}. 
Nevertheless, developing methods that alleviate the sign problem for a wide class of systems and extend the boundaries of current simulations is important. Ideally, this should be done in a controlled manner. 
Here, we propose a simple approach that provides accurate estimates of fermionic energies at lower temperatures. We demonstrate its usefulness for the model of electrons in two-dimensional QDs introduced in the previous Section, but the method can be more generally applied.
The key idea involves using the Bogoliubov inequality~\cite{feynman1972} to obtain an upper bound on the fermionic energies. 

\begin{figure}
\includegraphics[width=0.75\columnwidth]{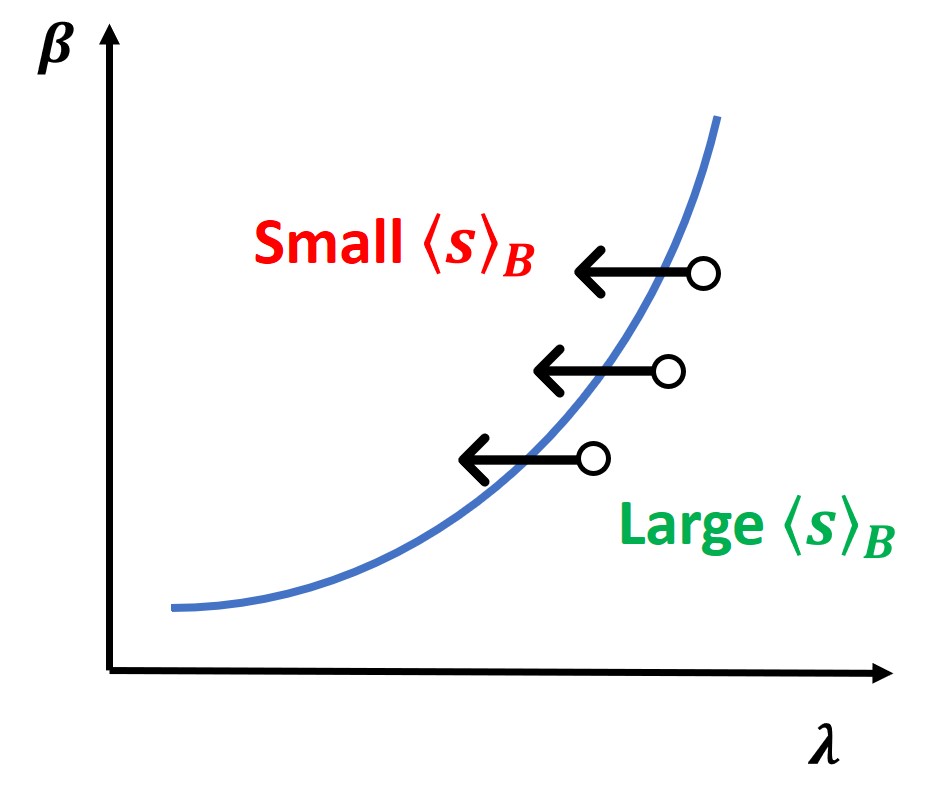}
\caption{\label{fig:sign} For a fixed number of particles, the average sign depends on the inverse temperature $\beta$ and the repulsion strength $\lambda$. The blue line represents some (arbitrary) critical value of the average sign. When the sign of the system is smaller (red), simulations are too expensive to converge. We propose simulating an auxiliary system that has a larger sign (green), and can be converged in practical simulations. Then, we estimate the energy of the original system, using the Bogoliubov inequality, at lower temperatures than otherwise would have been feasible using PIMD.}
\end{figure}

If the system Hamiltonian is denoted by $\hat{H}$ and the Hamiltonian of a second, auxiliary system is denoted by $\hat{H}'$, an upper bound on the difference in free energies between the two systems is given by the Bogoliubov inequality~\cite{feynman1972}

\begin{equation}
F_{\hat{H}} - F_{\hat{H}'} \le \left\langle \hat{H} - \hat{H}' \right\rangle_{\hat{H}'}.
\label{eq:Bogoliubov}
\end{equation}
In the following, we also assume that the energy of the system at low temperatures can be approximated by its free energy

\begin{equation}
\left\langle E \right\rangle_{\hat{H}} \approx \left\langle E \right\rangle_{\hat{H}'} + F_{\hat{H}} - F_{\hat{H}'}.
\label{eq:EisF}
\end{equation}
Eq.~\ref{eq:Bogoliubov} is very general and we can choose the auxiliary Hamiltonian such that it alleviates the sign problem. For example, it was shown in the previous Section that the average sign increases with repulsive interaction strength. Thus, it is possible to converge the simulations for strongly repelling fermions at temperatures lower than for weakly repelling ones. If we choose the former as our auxiliary Hamiltonian, we can obtain an upper bound on the energy of the weakly interacting system at low temperatures using the equations above. This is illustrated schematically in Fig.~\ref{fig:sign}. 

\begin{figure}
\includegraphics[width=1.0\columnwidth]{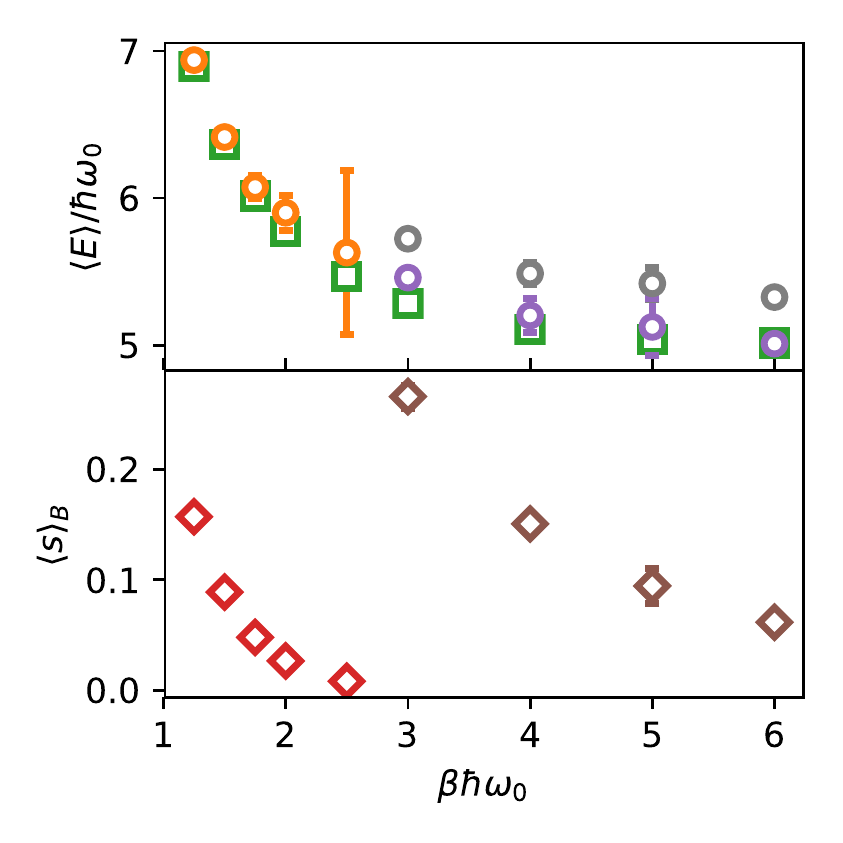}
\caption{\label{fig:3e} The average thermal energy (upper panel) and sign (lower panel) for three non-interacting electrons in a two-dimensional QD as a function of inverse temperature. Orange circles represent the results obtained by converging Eq.~\ref{eq:reW} directly. Gray circles represent the results obtained for an auxiliary system and purple circles show the results obtained using the Bogoliubov inequality. Analytical results are represented by green squares. If error bars are not shown, they are smaller than symbol size. In the lower panel, red and brown diamonds present the average sign of the non-interacting system and the auxiliary system, respectively.}
\end{figure}

We demonstrate this approach for a system of three electrons confined to a two-dimensional QD~\cite{Li2007} in the limit of infinite screening ($\lambda = 0$). As was shown in section~\ref{sec:QDs}, this leads to the most severe sign problem. This is also emphasized in the upper panel of Fig.~\ref{fig:3e}. Orange circles show that converging the simulations for the non-interacting system becomes difficult at $\beta=2.5$ where the error is visibly much larger ($\sim 10\%$) than at higher temperatures. 

Since it was shown by Dornheim~\cite{Dornheim2019} that short-range potentials have a larger average sign than systems with long-range interactions, we choose an auxiliary Hamiltonian in which the electrons interact with a repulsive Gaussian pair potential

\begin{equation}
V(|{\bf r}_l - {\bf r}_m|) = \frac{g}{\pi s^2}e^{-\frac{({\bf r}_l - {\bf r}_m)^2}{s^2}}.
\label{eq:int_pot}
\end{equation}
In Eq.~\ref{eq:int_pot}, $g$ and $s$ represent the interaction strength and range, respectively. In the following simulations, we used $g=16$ and $\sigma=0.5$ in harmonic oscillator units. The effective mass was set to $m=m_e$ and the frequency of the trap to $\hbar \omega_0 = 3 $ meV, as done by Runeson et al.~\cite{Runeson2018}. For each temperature, we performed 5 independent simulations of $75 \cdot 10^6$ steps each. Expectation values were converged with respect to the number of beads using $P=12$ for the non-interacting system and $P=72$ for the auxiliary system.

The bottom panel of Fig.~\ref{fig:3e} provides insight into why this approach is expected to succeed. The red symbols show the average sign of the non-interacting system as a function of temperature. It decreases sharply and is equal to $\approx 0.008$ at $\beta = 2.5$, which leads to the large error bars at this temperature in the upper panel. In contrast, the average sign for the auxiliary system decays significantly more slowly, which allows simulations at lower temperatures to be converged more easily.

The thermal energies obtained for the auxiliary system are shown in Fig.~\ref{fig:3e} (gray circles) and compared to the analytical results (green squares) for the non-interacting system. They are higher than the energies of non-interacting particles due to the added repulsion, as expected. Applying the Bogoliubov inequality (purple circles), we obtain upper bounds on the energies of the non-interacting system that agree with the analytical results within the statistical error. Most importantly, our approach allows an accurate estimate of the energy of the three-electron QD to be obtained at temperatures three times lower than would otherwise be feasible.  

 The results are not very sensitive to the value of the interaction parameter, as shown in the SI, but it must be large enough to converge the simulations at lower temperatures. We note, however, that this is likely to be  dependent on the type of repulsive interaction in the auxiliary system. In this case, the energy of the repulsive system grows slowly with the interaction parameter $g$ (see SI). In other cases, one should remember that the Bogoliubov inequality provides only an upper bound. Therefore, a good strategy is to choose a repulsion large enough to converge the simulations at low temperatures but as small as possible to obtain an upper bound which is close to the real expectation value. One could also imagine choosing variationally the auxiliary potential, which is beyond the scope of this work. 

\section{Conclusions}
\label{sec:conclusions}

In this Communication, we present a new method for performing PIMD simulations for fermions. Since direct sampling of fermionic partition functions is not feasible in PIMD (or PIMC), we perform simulations as if the system were composed of bosons and reweight the result to obtain fermionic expectation values using Eq.~\ref{eq:reW}. We apply the new method for up to $N=7$ electrons in a two-dimensional QD and obtain excellent agreement with PIMC results. However, like PIMC and other exact methods, our approach suffers from the fermionic sign problem. We show that the average sign decreases, as expected, with increasing $\beta$ and system size but also with decreasing repulsion strength. To alleviate the sign problem for weakly interacting systems, we propose performing simulations of an auxiliary system with a larger average sign (\textit{e.g.} by adding a fictitious repulsive interaction potential) and correcting its thermal energy using the Bogoliubov inequality. The results obtained in this way provide an upper bound to the energy of the weakly interacting systems. We show that using this approach we are able to solve the problem of a three-electron QD in the limit of infinite screening, for which the sign problem is most severe. We obtain accurate energies at temperatures that are three times lower than would have been possible by trying to converge the reweighting scheme directly. Challenges still remain in order to perform such simulations for a large class of interesting systems. For example, at even lower temperatures (or for larger systems) our method too will suffer from the sign problem. However, this work extends the boundaries of PIMD simulations of fermions and we are hopeful that it will encourage development of new methods to attack this important problem. 

\begin{acknowledgments}
This research was supported by the European Union Grant No. ERC-2014-ADG-670227/VARMET and the NCCR MARVEL, funded by the Swiss National Science Foundation. Calculations were carried out on the Euler cluster at ETH Zurich.
\end{acknowledgments}



%
%

%


\bibliography{BH_MI_MP}

\end{document}



\title{Supporting Information for ``Path Integral Molecular Dynamics for Fermions: Alleviating the Sign Problem with the Bogoliubov Inequality"} 



\author{Barak Hirshberg}
 \email{barakh@ethz.ch}
\affiliation{Department of Chemistry and Applied Biosciences, ETH Zurich, 8092 Zurich, Switzerland}
\affiliation{Institute of Computational Sciences, Universit\`{a} della Svizzera italiana, via G. Buffi 13, 6900 Lugano, Switzerland}%

\author{Michele Invernizzi}%
\affiliation{Institute of Computational Sciences, Universit\`{a} della Svizzera italiana, via G. Buffi 13, 6900 Lugano, Switzerland}%
\affiliation{National Centre for Computational Design and Discovery of Novel Materials MARVEL, Universit\`{a} della Svizzera italiana, via G. Buffi 13, 6900 Lugano, Switzerland}%
 \affiliation{Department of Physics, ETH Zurich, 8092 Zurich, Switzerland}

\author{Michele Parrinello}
 \affiliation{Department of Chemistry and Applied Biosciences, ETH Zurich, 8092 Zurich, Switzerland}
\affiliation{Institute of Computational Sciences, Universit\`{a} della Svizzera italiana, via G. Buffi 13, 6900 Lugano, Switzerland}%
\affiliation{Atomistic Simulations, Italian Institute of Technology, Via Morego 30, 16163 Genova, Italy}



\pacs{}

\maketitle 

\section{The virial estimator for bosons and fermions}
\label{sec:virial}

The quantum virial theorem is valid for both distinguishable and indistinguishable particles

\begin{equation}
    \langle \hat{T} \rangle_I = \frac{1}{2} \langle \sum_{l=1}^N {\bf \hat{r}}_l \cdot {\bf \hat{\nabla}}_{{\bf r}_l} V({\bf r}_1,...,{\bf r}_N) \rangle_I,
\end{equation}
where $I$ represents whether the particles are bosons (B), fermions (F) or distinguishable (D).
Therefore, the kinetic energy of bosons and fermions can be evaluated by using the estimator due to Berne et al.~\cite{Herman1982},

\begin{equation}
\varepsilon_T = \frac{1}{2P}\sum_{j=1}^P \sum_{l=1}^N {\bf r}_l^j \cdot {\bf \nabla}_{{\bf r}_l^j} V({\bf r}_1^j,...,{\bf r}_N^j).
\label{eq:Est}
\end{equation}
The only difference is that the expectation value is taken in the suitable bosonic or fermionic fictitious classical ensemble

\begin{equation}
\left\langle \hat{T} \right\rangle_I = \frac{1}{Z_I} \int \varepsilon_T e^{-\beta U^{(N)}_I} dR_1 ... dR_N \equiv \left\langle \varepsilon_T \right\rangle_I,
\label{eq:Av}
\end{equation}
where $Z_I$ and $ U^{(N)}_I$ are defined in the main text. 
It is well known that this estimator is not translationally invariant but for the trapped systems considered in this paper it is not an issue. As explained in detail in the main text, Eq.~\ref{eq:Av} is used directly to evaluate the bosonic expectation values  while fermionic averages are obtained from the bosonic simulations by reweighting.

\section{Statistical error estimation}

Care must be taken when evaluating statistical errors using weighted data since different simulations may not sample equally well the reweighted ensemble of interest. For example, if one simulation mostly explored configurations which have small weights it will provide a poor estimate of the reweighted energy. This is taken into account in the error estimation by associating the result obtained from each trajectory with a weight that is equal to the sum over all instantaneous weights during the simulation.

More specifically, the results presented in the main text were computed as follows. 
For each data point, we performed $M$ independent simulations of equal length. The first $10\cdot 10^6$ steps of each simulation were used for equilibration and were discarded from the statistical analysis. 
Then, we evaluated the fermionic energy of simulation $j$ using Eq. 9 of the main text, denoted as $E_j$. We also associated with each simulation a weight $W_j$ defined as the sum over all instantaneous weights $s=W_F^{(N)}/W_B^{(N)}$ in that trajectory.

The reported energies are the weighted average over all independent simulations

\begin{equation}
    \bar{E}_F = \frac {\sum_{j=1}^{M} E_j W_j}{\sum_{j=1}^{M} W_j}.
\end{equation}
Then, we evaluated the effective sample size for our data,
\begin{equation}
    n = \frac{(\sum_{j=1}^M W_j)^2}{\sum_{j=1}^M W_j^2}.
\end{equation}
In our simulations, the effective sample size was close to the number of trajectories M, indicating that they were all similarly converged.
An unbiased estimate of the variance~\cite{Ambegaokar2010} is then evaluated using

\begin{equation}
    \sigma_E^2 = \frac{n}{n-1} \frac{\sum_{j=1}^{M} W_j (E_j - \bar{E}_F)^2 }{\sum_{j=1}^{M} W_j},
\end{equation}
and the statistical error associated with $\bar{E}_F$ is given by $\sigma_E/\sqrt{n}$. 

We note that when $W_j=1$ the equations above reduce to the simple average of non-weighted data, which is used for evaluating the bosonic energies and the average sign $\langle s \rangle_B$. In that case, the value for each independent simulation is obtained directly from the average of the appropriate estimator without rewighting.

\section{Sensitivity to interaction strength}
\begin{figure}
\includegraphics[width=0.5\columnwidth]{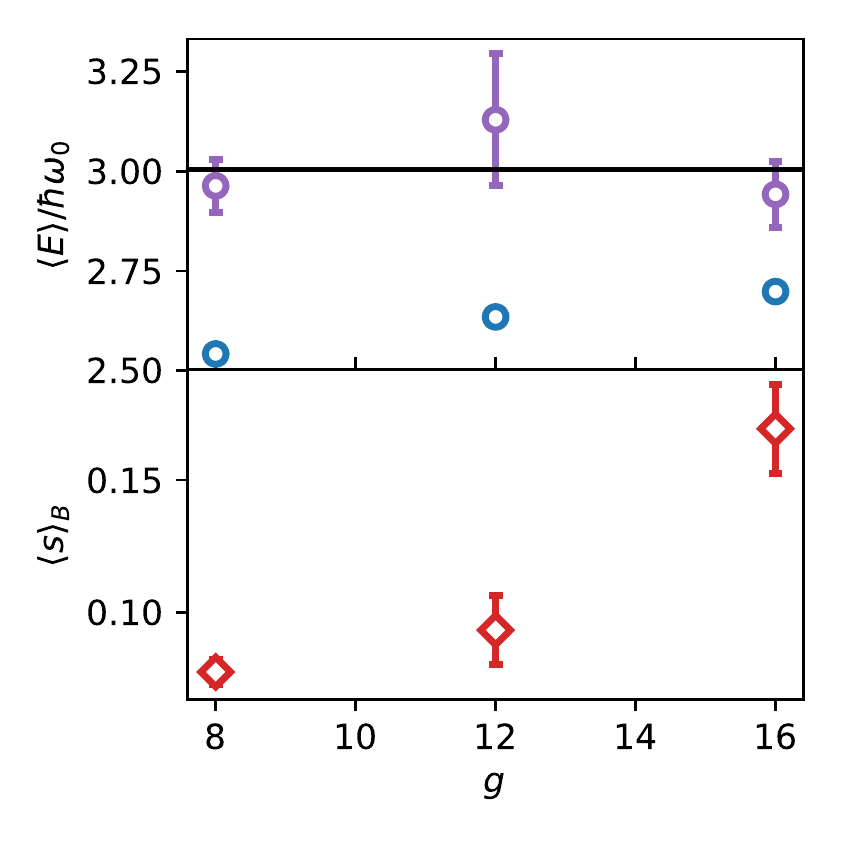}
\caption{\label{fig:Evsg} The average thermal energy as a function of the interaction strength $g$ for two particles. The inverse temperature is $\beta \hbar \omega_0 = 6$. Blue circles represent the energies of the auxiliary bosonic system. Purple circles show the fermionic energies of the non-interacting system obtained using the Bogoliubov inequality. The black line is the analytical result. If not shown, error bars are smaller than symbol size.}
\end{figure}

We also performed simulations for a two-electron quantum dot (QD) in the infinite screening limit, using the same auxiliary Hamiltonian as in the main text, and tested the sensitivity of the results to the interaction strength parameter $g$.
The upper panel of Fig.~\ref{fig:Evsg} shows the energy of the auxiliary system as a function of $g$. It can be seen that the bosonic energies (blue) grow relatively slowly with the interaction parameter. As a result, the fermionic energies of the non-interacting system obtained from the Bogoliubov inequality are not very sensitive to the value of $g$ (purple). The lower panel shows the average sign as a function of interaction strength.

%
%

%


\bibliography{BH_MI_MP}